# Theoretical investigation of power absorbed by superconductors with Rashba spin-orbit coupling in the presence external time-dependent perturbation

**Reza Afzali[1], Bahman Takrim[1], Amir Hossein Asghari[1]**

[1] Department of Physics, K.N.Toosi University of Technology, Tehran 15875-4416 Iran

**Abstract**

The absorbed power in superconductors driven by a time-dependent external perturbation is investigated through transitions between Bogoliubov quasiparticle states. The transition matrix element contains the superconducting coherence factor, which determines the role of the gap, excitation spectrum, and external frequency in the absorption process. The formulation is applied to a Hubbard-model superconductor including correlated hopping and Rashba spin-orbit coupling. In this model, numerical calculations are used to obtain the quasiparticle energies, density of states, chemical potential, and temperature-dependent superconducting gap. The normalized absorbed power is then evaluated for different values of temperature, interaction strength, correlated hopping parameter, and spin-orbit coupling strength. The results show that the absorption response is strongly affected by the superconducting gap and by changes in the quasiparticle spectrum. Rashba spin-orbit coupling changes the density of states and therefore modifies the absorption. These results show that hopping processes, interaction strength, and spin-orbit coupling play important roles in the ultrasonic and electromagnetic absorption of superconducting systems.

## 1 Introduction

More recent work by Nam in 2006 studied absorption in unconventional d-wave superconductors, revealing the role of the gap symmetry [1]. Absorption also occurs in the presence of spin-orbit coupling, as examined by Bergeret and Tokatly in 2013 [2]. The application of time-dependent electromagnetic perturbations to superconducting materials enables the study of their exotic properties and opportunities to further enhance real-world applications [3]. The absorbed power provides insights into the underlying quantum mechanical structure and dynamics in these remarkable states of matter [4].

In parallel, spin-orbit coupling was being investigated as a relativistic quantum effect linking electron spin and momentum [5]. The role of spin-orbit interactions in superconductors was largely unexplored until recently [6]. In 2008, Gor'kov and Rashba showed that spin-orbit coupling can induce mixed singlet-triplet pairing in superconductors [7]. Bergeret and Tokatly in 2013 found that spin-orbit coupling enables long-range triplet proximity effects [2]. External electromagnetic fields can strongly influence superconductors [4]. In 1970, Eliashberg studied microwave absorption in superconductors [8]. Nam in 2006 analyzed microwave absorption in d-wave superconductors [9]. Spin-orbit coupling alters the nature of superconductivity, while time-varying fields provide insights into exotic properties [10]. Advances in quantum engineering and 2D materials system open up new possibilities for exploiting spin-orbit physics to enhance superconductors [11]. The discovery of superconductivity marked a major scientific advancement. [12] Conventional s-wave superconductors exhibit an isotropic energy gap [13]. When subjected to electromagnetic radiation, they display absorption phenomena like the breaking of Cooper pairs [14]. Unconventional d-wave superconductors have an anisotropic gap function [15]. Nam showed that this angular dependence reduces absorption of microwaves in the superconducting state of d-wave superconductors [16].

Tight binding models have also been developed to study superconductivity in materials [17]. Alidoust et al. calculated the absorption power for a tight binding superconductor with and without spin-orbit coupling when exposed to a.c. electromagnetic fields [18]. The presence of spin-orbit interactions was found to impact absorption. Different types of superconductors display distinct behaviors in absorbing electromagnetic radiation [19]. The anisotropic gap function of d-wave superconductors reduces absorption [20]. Spin-orbit coupling alters absorption in tight binding model superconductors [6].we will investigate the power absorption phenomena in conventional tight-binding model superconductor both with and by considering spin-orbit coupling effect. Through numerical simulations and analysis, we hope to elucidate the role of spin-orbit coupling and superconducting gap symmetry in altering the electromagnetic absorption properties. The outcomes of this research will further the fundamental understanding of spin-orbit effects and absorbed power mechanisms in different superconducting systems.

**Transition probabilities**

In superconductors, the absorbed power resulting from time-dependent external perturbations is an important phenomenon that can be studied using probability transition computations and Bogoliubov transformations. When an external perturbation, such as Microwave perturbations, is applied to a superconductor, it can induce transitions between the superconducting ground state and excited states. The probability of these transitions occurring can be calculated using time-dependent perturbation theory. Bogoliubov transformations, which mix electrons and holes, can then be applied to determine the populations of the excited states. Knowing the transition rates and populations allows the power absorbed by the superconductor from the external perturbation to be calculated. The perturbation induced in the system can be expressed by:

$$H=\sum_{k,k',\alpha,\alpha'} B(\vec{k}\alpha|\vec{k}'\alpha')[u_k u_{k'}\gamma^*_{k\beta}\gamma_{k'\beta'}+v_k v_{k'}\sum_{\beta,\beta'}\rho_{\alpha\beta}\rho_{\alpha'\beta'}\gamma_{k\beta}\gamma^*_{k'\beta}$$

$$+u_k v_{k'}\sum_{\beta}\rho_{\alpha'\beta'}\gamma^*_{k\beta}\gamma^*_{-k'\beta'}+u_{k'}v_k\sum_{\beta}\rho_{\alpha\beta}\gamma_{-k\beta}\gamma_{-k'\beta'}$$

The matrix element $\mathrm{M}(\vec{k}\alpha|\vec{k}'\alpha')$ represents the coefficient of $\gamma^*_{k\beta}\gamma_{k'\beta'}$ that relates the two states in the transition and given by:

$$M(\vec{k}\alpha|\vec{k}'\alpha')=B(\vec{k}\alpha|\vec{k}'\alpha')u_k u_{k'}-v_k v_{k'}\sum_{\sigma,\sigma'}\rho_{\sigma'\alpha'}\rho_{\sigma\alpha}B(-\vec{k}'\alpha'|-\vec{k}\alpha)$$

Where the expression $\sum_{\sigma,\sigma'}\rho_{\sigma'\alpha'}\rho_{\sigma\alpha}B(-\vec{k}'\alpha'|-\vec{k}\alpha)$ represents the matrix elements B with the spins and momenta of the electrons conjugated. For the interaction $H$ that we had assumed, the expression $B(\vec{k}\alpha|\vec{k}'\alpha')$ is denoted as:

$$\sum_{\sigma,\sigma'}\rho_{\sigma'\alpha'}\rho_{\sigma\alpha}B(-\vec{k}'\alpha'|-\vec{k}\alpha)=\eta B(\vec{k}\alpha|\vec{k}'\alpha')$$

And then:

$$M(\vec{k}\alpha|\vec{k}'\alpha')=B(\vec{k}\alpha|\vec{k}'\alpha')\left[u_k u_{k'}-\eta v_k v_{k'}\right]$$

Where the expression $\left[u_k u_{k'}-\eta v_k v_{k'}\right]$ is called the coherence factor and $\vec{k}\alpha,\vec{k}'\alpha'$ represent the probability of the electron existing in the two states $u$ and $v$ with the wave vector $\vec{k}$ respectively and the expression is as follows:

$$|v_k|^2 = 1 - |u_k|^2 = \frac{1}{2}\left(1 - \frac{\xi_k}{\varepsilon_k}\right)$$

In which the parameter ξ represents the coherence length of a superconductor and by using above equation, we obtain:

$$[u_k u_{k'} - \eta v_k v_{k'}] = \frac{1}{2}\left[1 - \eta \frac{\Delta^2}{\varepsilon\varepsilon'}\right]$$

And $\eta$ is a constant parameter which is equal +1 for case I processes (Ultrasound absorption) and -1 for case II processes (Electromagnetic absorption).

The transition rate from $\vec{k}\alpha$ to $\vec{k}'\alpha'$ state for the Bogoliubov quasiparticles is given by:

$$\dot{v} = \frac{2\pi}{\hbar} |M(\vec{k}\alpha|\vec{k}'\alpha')|^2 \left[ f(\varepsilon_{k'})\left[1 - f(\varepsilon_k)\right] - f(\varepsilon_k)\left[1 - f(\varepsilon_{k'})\right]\delta(\varepsilon_k - \varepsilon_{k'} - \hbar\omega)\right.$$

Where the term $M(\vec{k}\alpha|\vec{k}'\alpha')$ represents the matrix elements associated with the quasiparticle transitions and $f(\varepsilon_k)$ is the Fermi-Dirac distribution function for the electrons. We have assumed that $H$, is a sinusoidal perturbation of frequency $\omega$, The Absorbed power is given by:

$$W = \sum_{k.k'} \dot{v}\hbar\omega$$

By substitutingin we will have: $\dot{v}$

$$W = \sum_{k,k'} \frac{2\pi}{\hbar}[B(\vec{k}\alpha|\vec{k}'\alpha')(u_k u_{k'} - \eta v_k v_{k'})]^2 [f(\varepsilon_{k'})1 - f(\varepsilon_{k'})]\delta(\varepsilon_{k'} - \varepsilon_k - \hbar\omega)$$

In which $\varepsilon_k$ is the kinetic energy of quasiparticles. To calculate $W$ we first need to average over the angles of $\vec{k}$ and $\vec{k}'$ as well as the spin indices of the matrix element $B$:

$$B = \sqrt{B(\vec{k}\alpha|\vec{k}'\alpha')^2}$$

In the above relation, $B$ is treated as a constant and the absorbed power will be:

$$W = 2\pi\omega B^2 N(0)^2 \int_{-\infty}^{\infty}\int_{-\infty}^{\infty} g(\varepsilon)g(\varepsilon')[uu' - \eta vv']^2 \left[ f\left(\varepsilon'\right) - f\left(\varepsilon\right)\right] \delta\left(\varepsilon - \varepsilon' - \hbar\omega\right)] d\varepsilon d\varepsilon'$$

In which $g(\varepsilon)$ is density state of quasiparticles and $N(0)$ is the normal state density. and $\omega$ is the frequency that manifests due to an external perturbation, $f(E_{ks}) = \frac{1}{1+e^{\frac{E_{ks}-\mu}{k_B T}}}$ and $f(E_{ks}) = \frac{1}{1+e^{\frac{E_{k's'}-\mu}{k_B T}}}$ are the Fermi-Dirac distribution functions in which $k_B$ is the Boltzmann Constant and $T$ is the system temperature. Moreover, the characteristic $B$ corresponds to the matrix interaction elements and is considered as constant parameters.

**Measuring Absorbed Power for a Hubbard Model Superconductor in the Presence of Spin-Orbit Effect**

The Hubbard model, first proposed by John Hubbard in 1963, is a simplified model that aims to describe the electronic properties of some superconducting materials. It considers a lattice of sites that can each be occupied by at most one electron. The model includes two key terms: a kinetic energy term representing electron hopping between sites, and an on-site repulsion term that accounts for the Coulomb repulsion when two electrons occupy the same site. By tuning the relative strengths of these terms, the Hubbard model can generate different many-body phenomena. Of particular interest is the prediction of d-wave superconductivity when the on-site repulsion U is comparable to the electron hopping kinetic energy t. Solving the Hubbard model has provided insights into the mechanism of high-temperature superconductivity observed in cuprite materials, where electron-electron interactions are believed to mediate the formation of Cooper pairs responsible for superconductivity. The Hubbard model demonstrates how complex strongly-correlated electron behaviors can emerge from a simple theoretical construct [21]. By following the standard BCS method, we Fourier transform the tight-binding Hamiltonian that includes correlated hopping and spin-orbit coupling terms. To simplify the analysis, we then remove interactions between pairs of particles with nonzero momentum. This allows us to obtain a simplified Hamiltonian that we can analyze further for the Hubbard model system [22]:

$$H = \sum_{k,\sigma}(\varepsilon_k - \mu)c^\dagger_{k\sigma}c_{k\sigma-} 2V_{so}\sum_k (sink_y(c^\dagger_{k\uparrow}c_{k\downarrow} + c^\dagger_{k\downarrow}c_{k\uparrow})$$
$$+isink_x(c^\dagger_{k\uparrow}c_{k\downarrow} - c^\dagger_{k\downarrow}c_{k\uparrow}) + \frac{1}{N}\sum_k V^0(k,k')c^\dagger_{k\uparrow}c^\dagger_{-k\downarrow}c_{-k'\downarrow}c_{k'\uparrow}$$
$$+\frac{1}{N}\sum_{KK'}\sum_{\alpha\beta}(V^R_{\alpha\beta}(K')c^\dagger_{k\alpha}c^\dagger_{-k\beta}c_{k'\alpha}c_{k'\alpha} + V^R_{\alpha\beta}(K)c^\dagger_{k\beta}c^\dagger_{-k\beta}c_{k'\beta}c_{-k'\alpha}$$

Here $N$ is the number of $K$ points, the operators $c_{k\sigma}$ creates an electron on site $(k)$ with spin $(\sigma = \uparrow, \downarrow)$ within the tight-binding model. The Rashba spin-orbit coupling strength is represented by the parameter $V_{so}$ and when $V_{so} = 0$ , indicating the Rashba interaction is absent, the dispersions and interactions of the tight-binding electronic states are determined solely by:

$$\sum_k V^0(k,k') \equiv U - 2\frac{\Delta t}{t}(\varepsilon_k + \varepsilon_{k'})$$

When spin-orbit coupling is not taken into account. Diagonalzing the non-interacting Hamiltonian analytically determines the spin-orbit coupled spectrum before examining the interacting system. We consider coupling of hopping and Rashba terms by

$$V^R_{\alpha\beta}(k) \equiv 2V_{so}2\frac{\Delta t}{t}(\sin k_x\sigma^y_{\alpha\beta} + \sin k_y\sigma^x_{\alpha\beta})$$

In which $U$ and $\Delta t$ represent the on-site repulsion and the correlated hopping parameter respectively. The tight-binding model defines the electron hopping between lattice sites using the hopping integral parameter $t$, and the wave vectors along the $x$ and $y$ directions are denoted as $k_x$ and $k_y$ [23]. We set the lattice constant to unity and diagonalzing the non-interacting Hamiltonian in the Rashba basis to enable determining spectrum:

$$\varepsilon_{ks} = \varepsilon_k - 2sV_{so}\sqrt{\sin^2 k_x + \sin^2 k_y}$$

The variable $s$ can take on the values $+1$ or $-1$, representing two different helicity states. The kinetic and excitation energy relations of electrons in the Hubbard in presence of hopping and spin-orbit effect, described by:

$$\begin{cases} \varepsilon_k = -2t(\cos k_x + \cos k_y) \\ E_{ks} = \sqrt{(\varepsilon_{ks} - \mu)^2 + \Delta_{ks}{}^2} \end{cases}$$

Where $\mu$ is the chemical potential, $\varepsilon_{ks}$ is the quasiparticle kinetic energy, and $\Delta_{ks}$ denotes the superconducting energy gap, which is described by the following:

$$\begin{cases} \Delta_{ks} = \Delta^0 - \Delta^s \dfrac{\varepsilon_{ks}}{4t} \\ \Delta_{k's'} = \Delta_{k's'} = \Delta^0 - \Delta^{s'} \dfrac{\varepsilon_{k's'}}{4t} \end{cases}$$

To determine the absorbed power $(W)$ within the Hubbard model framework, we first derive expressions representing all relevant relations with respect to the electron excitation energies. The excitation properties serve to define the system's quantized transitional states that mediate the absorption and dissipation of energy. We establish these governing relations in terms of the excitation spectrum. The variables $\Delta^0$ and $\Delta^s$ are temperature-dependent parameters independent of the excitation energies in the system (see the section 2.2). Rather, these quantities are purely functions of thermal effects, decoupled from the electron excitation spectrum. Hence we express $\varepsilon_{ks}$ in terms of $E_{ks}$ as follows:

$$E_{ks} = \sqrt{(\varepsilon_{ks} - \mu)^2 + \Delta_{ks}{}^2} \Rightarrow (\varepsilon_{ks} - \mu)^2 = E_{ks}{}^2 - \Delta_{ks}{}^2$$

And then we get:

$$(\varepsilon_{ks} - \mu)^2 = E_{ks}{}^2 - \Delta_{ks}{}^2 \Rightarrow \varepsilon_{ks} - \mu = \sqrt{E_{ks}{}^2 - \Delta_{ks}{}^2} \Rightarrow \begin{cases} \varepsilon_{ks} = \sqrt{E_{ks}{}^2 - \Delta_{ks}{}^2} + \mu \\ \varepsilon_{k's'} = \sqrt{E_{k's'}{}^2 - \Delta_{k's'}{}^2} + \mu \end{cases}$$

To calculate the absorption power, the density of states is defined as in equation:

$$g(E) = 4\sum_s \int_0^\pi \frac{dk_x}{2\pi} \int_0^\pi \frac{dk_y}{2\pi} \delta\left(E_{ks} - \varepsilon_{ks}\right) = \lim_{\sigma \to 0} \frac{1}{\pi^{3/2}\sigma t} \sum_s \int_0^\pi dk_x \int_0^\pi dk_y e^{-(E/t - \varepsilon_{ks}/t)2} / \sigma^2$$

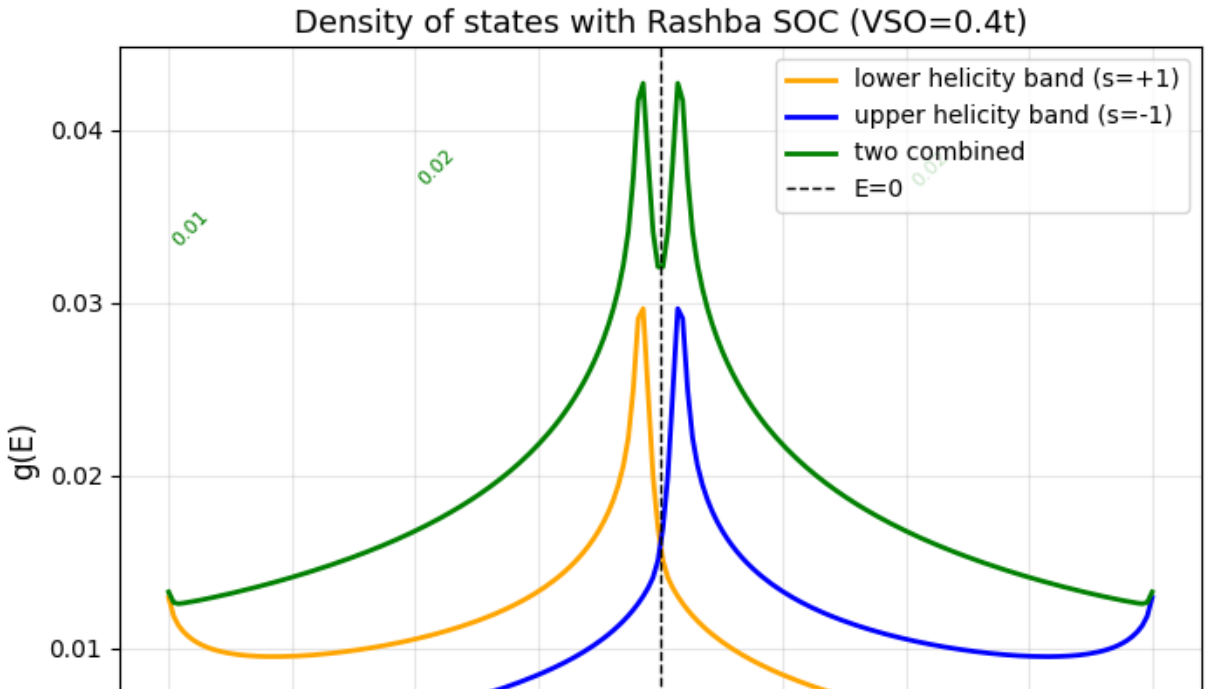

**FIG. 1.** The density of states function ( $g(E)$ ) as a function of ( $E/t$ ), obtained numerically using Python.

to obtain the temperature-dependent energy gap function, equations:

$$\overline{\Delta}_{ks} = \Delta^0 + \Delta^s (s_k + \frac{sV_{so}}{2t}\sqrt{\sin^2 k_x + \sin^2 k_y}) = \Delta^0 - \Delta^s \frac{\varepsilon_{ks}}{4t}$$

In which:

$$\Delta^0 = -\frac{1}{2N}\sum_{k's'}(U + 8\Delta t s_{k'} + 4\frac{\Delta t}{t}V_{so}s'\sqrt{\sin^2 k_x' + \cos^2 k_y'})g(E_{k's'})\overline{\Delta}_{k's'}$$

$$\Delta^s = -\frac{1}{2N}\sum_{k's'} 8\Delta t g(E_{k's'})\overline{\Delta}_{k's'}$$

and equations:

$$\eta = 1 - \frac{1}{2N}\sum_{k's'}\frac{(\varepsilon_{k's'} - \mu)}{E_{k's'}}\tanh(\beta E_{k's'}/2)$$

are solved simultaneously to produce the graph of the density of states versus energy. For this purpose, we use a numerical solution of the density of states function.

Based on the obtained results and data, we define the energy gap as a function of temperature, which is given by equation:

$$\Delta_{ks}(T) = \Delta_0 \sqrt{1 - (\frac{T}{T_c})^2}$$

To obtain the expression for the absorption power in the Hubbard model, we first rewrite the general relation for the absorption power as follows:

$$W = 2\pi\omega B^2 \int_{-\infty}^{\infty}\int_{-\infty}^{\infty} g(\varepsilon)g(\varepsilon')[uu' - \eta vv']^2 \left[ f\left(\varepsilon'\right) - f\left(\varepsilon\right)\right] \delta\left(\varepsilon - \varepsilon' - \hbar\omega\right)] d\varepsilon d\varepsilon'$$

Then we should obtain the $[uu' - \eta vv']^2$ factor:

$$[u_{ks}u'_{ks} - \eta v_{ks}v'_{ks}]^2 = \left( \sqrt{\frac{1}{2}\left(1 + \frac{\varepsilon_{ks} - u}{E_{ks}}\right)} . \sqrt{\frac{1}{2}\left(1 + \frac{\varepsilon'_{ks} - u}{E'_{ks}}\right)} - \sqrt{\frac{1}{2}\left(1 - \frac{\varepsilon_{ks} - u}{E_{ks}}\right)} . \sqrt{\frac{1}{2}\left(1 - \frac{\varepsilon_{ks} - u}{E_{ks}}\right)} \right)^2$$

Where $u$ the chemical potential, $\varepsilon_{ks}$ is the quasiparticle kinetic energy, $E_{ks}$ is the quasiparticle excitation energy and given by:

$$E_{ks} = \sqrt{(\varepsilon_{ks} - \mu)^2 + \Delta_{ks}{}^2}$$

We raise the resulting expression to the power of 2, then multiply the expressions inside the parentheses:

$$[u_{ks}u'_{ks} - \eta v_{ks}v'_{ks}]^2 = \frac{1}{4}[\left(1 + \frac{\varepsilon_{ks} - u}{E_{ks}} + \frac{\varepsilon'_{ks} - u}{E'_{ks}} + \frac{(\varepsilon_{ks} - u)(\varepsilon'_{ks} - u)}{E_{ks}E'_{ks}}\right) + (1 - \frac{\varepsilon_{ks} - u}{E_{ks}} - \frac{\varepsilon'_{ks} - u}{E'_{ks}}$$

$$+ \frac{(\varepsilon_{ks} - u)(\varepsilon'_{ks} - u)}{E_{ks}E'_{ks}} - \sqrt{\left(1 - \frac{(\varepsilon_{ks} - u)^2}{E_{ks}}\right)\left(1 - \frac{(\varepsilon'_{ks} - u)^2}{E'_{ks}}\right)} \;]$$

And then we get:

$$[u_{ks}u'_{ks}-\eta v_{ks}v'_{ks}]^2=\frac{1}{2}\left(1+\frac{(\varepsilon_{ks}-u)(\varepsilon'_{ks}-u)}{E_{ks}E'_{ks}}-\frac{\Delta_{ks}\Delta'_{ks}}{E_{ks}E'_{ks}}\right)$$

In the above relation, since the term $(\varepsilon_{ks}-u)$ is not an even function of $E_{ks}$ $\frac{(\varepsilon_{ks}-u)(\varepsilon'_{ks}-u)}{E_{ks}E'_{ks}}$ cannot be omitted. Similar to what we did to obtain the correlation factor for the scattering process, we can write for the quasiparticle creation and destruction processes. Then we get:

$$W=2\pi\omega B^2\int_{-\infty}^{\infty}\int_{-\infty}^{\infty}g(E_{ks})g(E_{k's'})\left[1+\frac{(\varepsilon_{ks}-u)(\varepsilon'_{ks}-u)}{E_{ks}E_{k's'}}+\eta\frac{\Delta_{ks}\Delta'_{ks}}{E_{ks}E_{k's'}}\right]$$

$$\left[f(E_{k's'})-f(E_{ks})\right]\delta(E_{ks}-E_{k's'}-\hbar\omega)]dE_{ks}dE_{k's'}$$

We can write the relation in term of $E$:

$$W=2\pi\omega B^2\int_{-\infty}^{\infty}\int_{-\infty}^{\infty}g(E_{ks})g(E_{k's'})\left[1+\frac{\left(\sqrt{E_{ks}{}^2-\Delta_{ks}{}^2}\right)\left(\sqrt{E_{k's'}{}^2-\Delta_{k's'}{}^2}\right)}{E_{ks}E_{k's'}}+\eta\frac{\Delta_{ks}\Delta_{k's'}}{E_{ks}E_{k's'}}\right]$$

$$\left[f(E_{k's'})-f(E_{ks})\right]\delta(E_{ks}-E_{k's'}-\hbar\omega)]dE_{ks}dE_{k's'}$$

Since obtaining the exact value of parameter $B$ is very difficult, to calculate the absorption power, we determine the ratio of the absorption power in the superconducting state to that in the normal state. To do this, it is sufficient to set the gap function equal to zero; as a result, we have:

$$\Delta_{ks}=0\Rightarrow W=W_N$$

$$\Rightarrow W_N=4\pi\omega B^2\int_{-\infty}^{\infty}\int_{-\infty}^{\infty}g(E_{ks})g(E_{k's'})\left[f(E_{k's'})-f(E_{ks})\right]\delta(E_{ks}-E_{k's'}-\hbar\omega)]dE_{ks}dE_{k's'}$$

The we can get:

$$\frac{W}{W_N}=\frac{\int\limits_{-\infty}^{\infty}\int\limits_{-\infty}^{\infty} g(E_{ks})g(E_{k's'})\left[1+\frac{\left(\sqrt{E_{ks}{}^{2}-\Delta_{ks}{}^{2}}\right)\left(\sqrt{E_{k's'}{}^{2}-\Delta_{k's'}{}^{2}}\right)}{E_{ks}E_{k's'}}+\eta\frac{\Delta_{ks}\Delta_{k's'}}{E_{ks}E_{k's'}}\right]\left[f\left(E_{k's'}\right)-f\left(E_{ks}\right)\right]\delta\left(E_{ks}-E_{k's'}-\hbar\omega\right)]dE_{ks}dE_{k's'}}{\int\limits_{-\infty}^{\infty}\int\limits_{-\infty}^{\infty} g(E_{ks})g(E_{k's'})\left[f\left(E_{k's'}\right)-f\left(E_{ks}\right)\right]\delta\left(E_{ks}-E_{k's'}-\hbar\omega\right)]dE_{ks}dE_{k's'}}$$

# Results

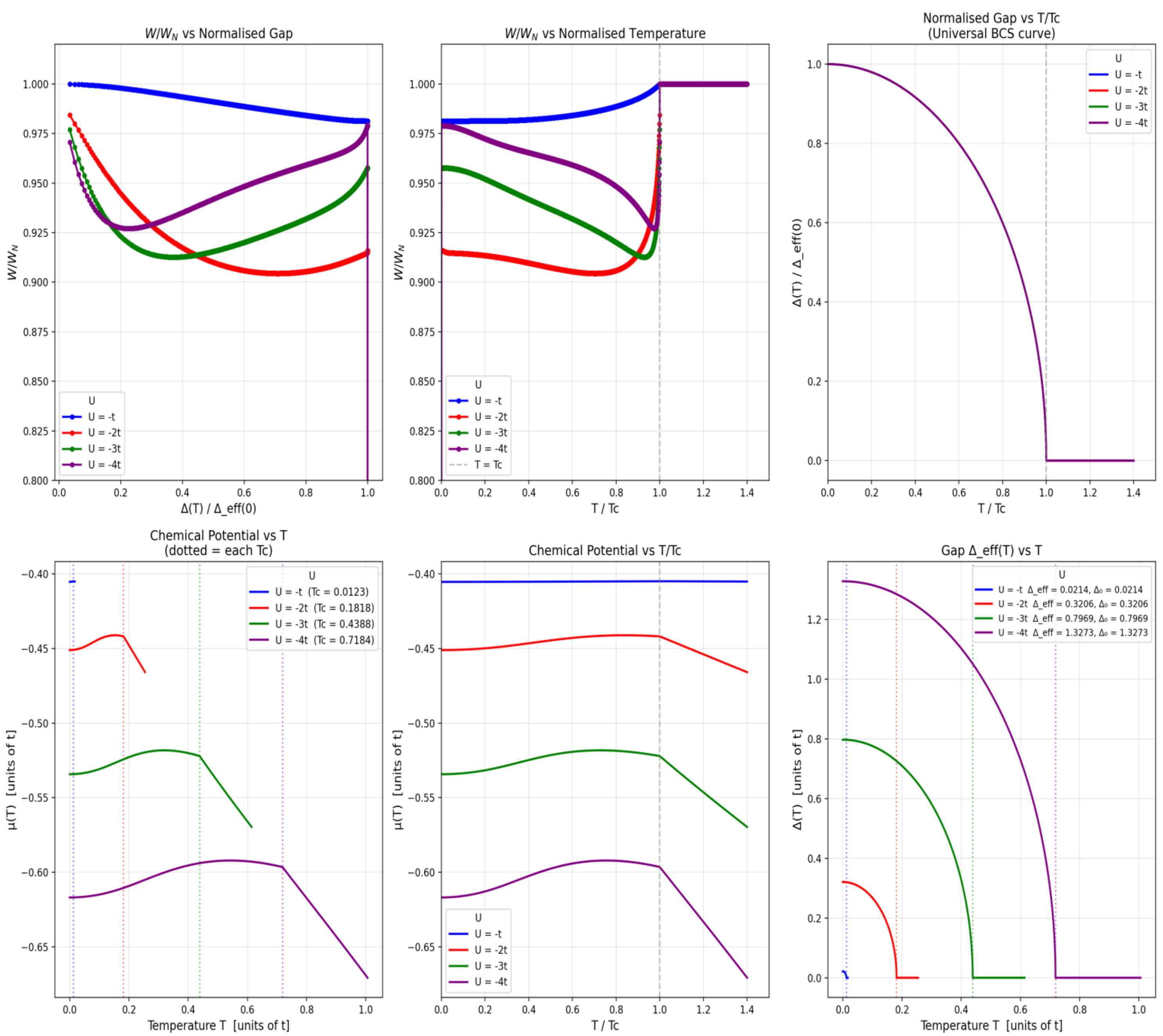

Ultrasonic Attenuation — Group B (U = 75t...125t, Δ_t = 4, n = 1.8850, V_SO = 0)

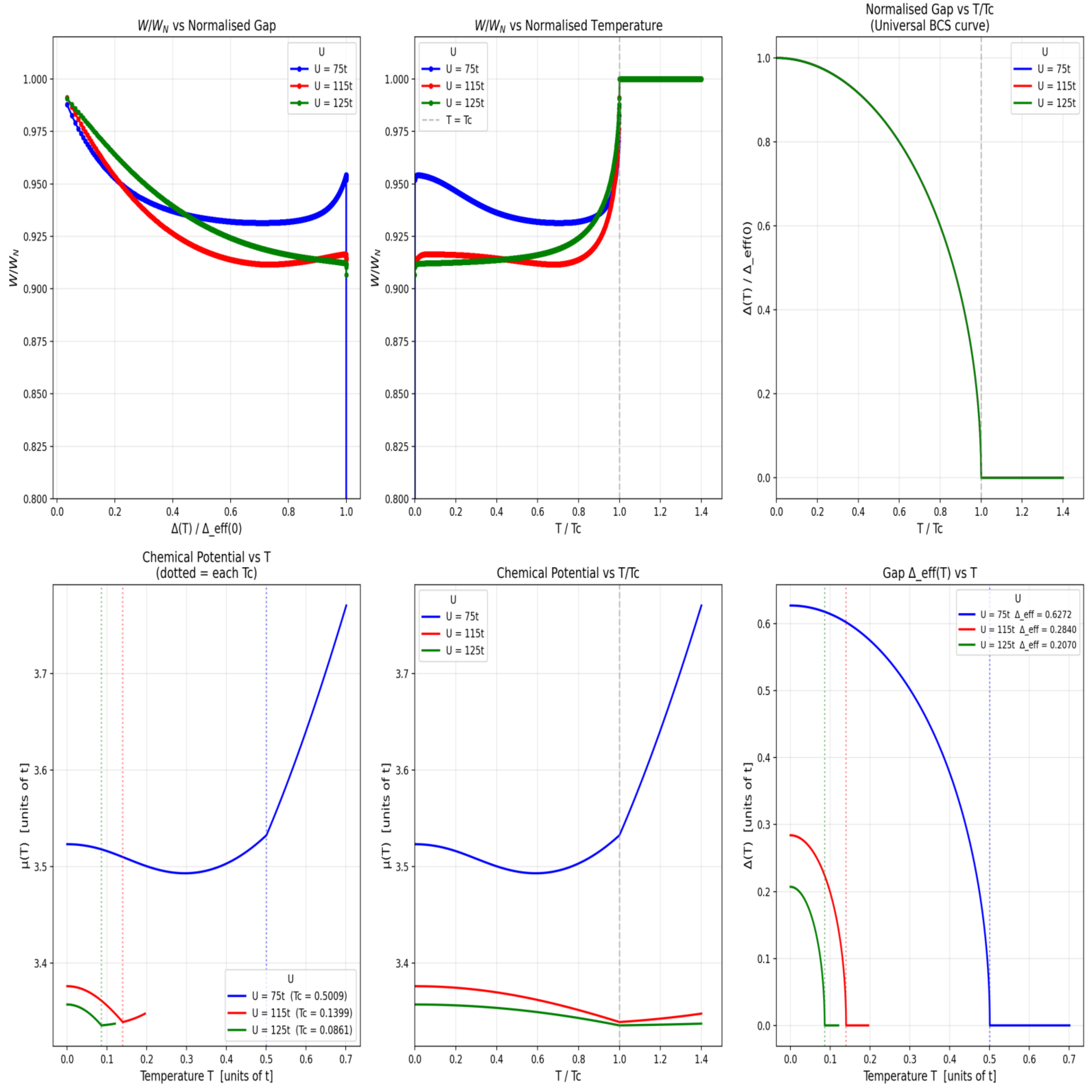

Ultrasonic Attenuation — Group C (U = 75t, n = 1.8850, V_SO = 0.8, Δ_t varies)

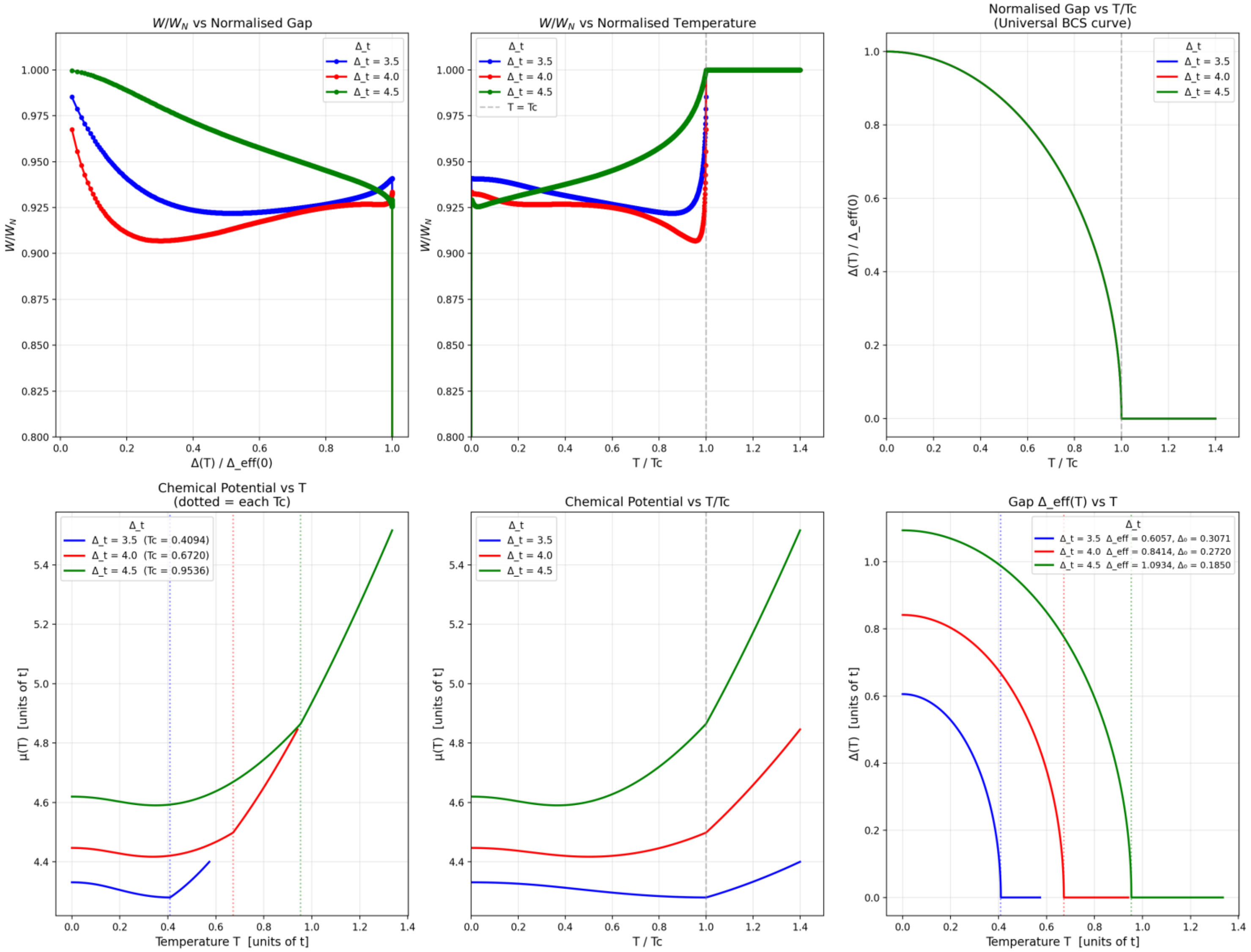

$W/W_N$ vs Normalized Gap

VSO = 0.0, VSO = 0.4, VSO = 0.8

$W/W_N$ — $\Delta(T)/\Delta_0$

$W/W_N$ vs Normalized Temperature

VSO = 0.0, VSO = 0.4, VSO = 0.8, T = Tc

$W/W_N$ — T / Tc

Normalized Gap vs Normalized Temperature
(Should be universal BCS curve)

VSO = 0.0, VSO = 0.4, VSO = 0.8

$\Delta(T)/\Delta_0$ — T / Tc

Chemical Potential vs Temperature
(Vertical lines mark each Tc)

VSO = 0.0 (Tc=0.718), VSO = 0.4 (Tc=0.686), VSO = 0.8 (Tc=0.590)

μ(T) — Temperature T

Chemical Potential vs Normalized Temperature

VSO = 0.0, VSO = 0.4, VSO = 0.8

μ(T) — T / Tc

Gap vs Temperature
(Different Tc for each VSO)

VSO=0.0 (Tc=0.718, $\Delta_0$=1.327), VSO=0.4 (Tc=0.686, $\Delta_0$=1.262), VSO=0.8 (Tc=0.590, $\Delta_0$=1.072)

Δ(T) — Temperature T

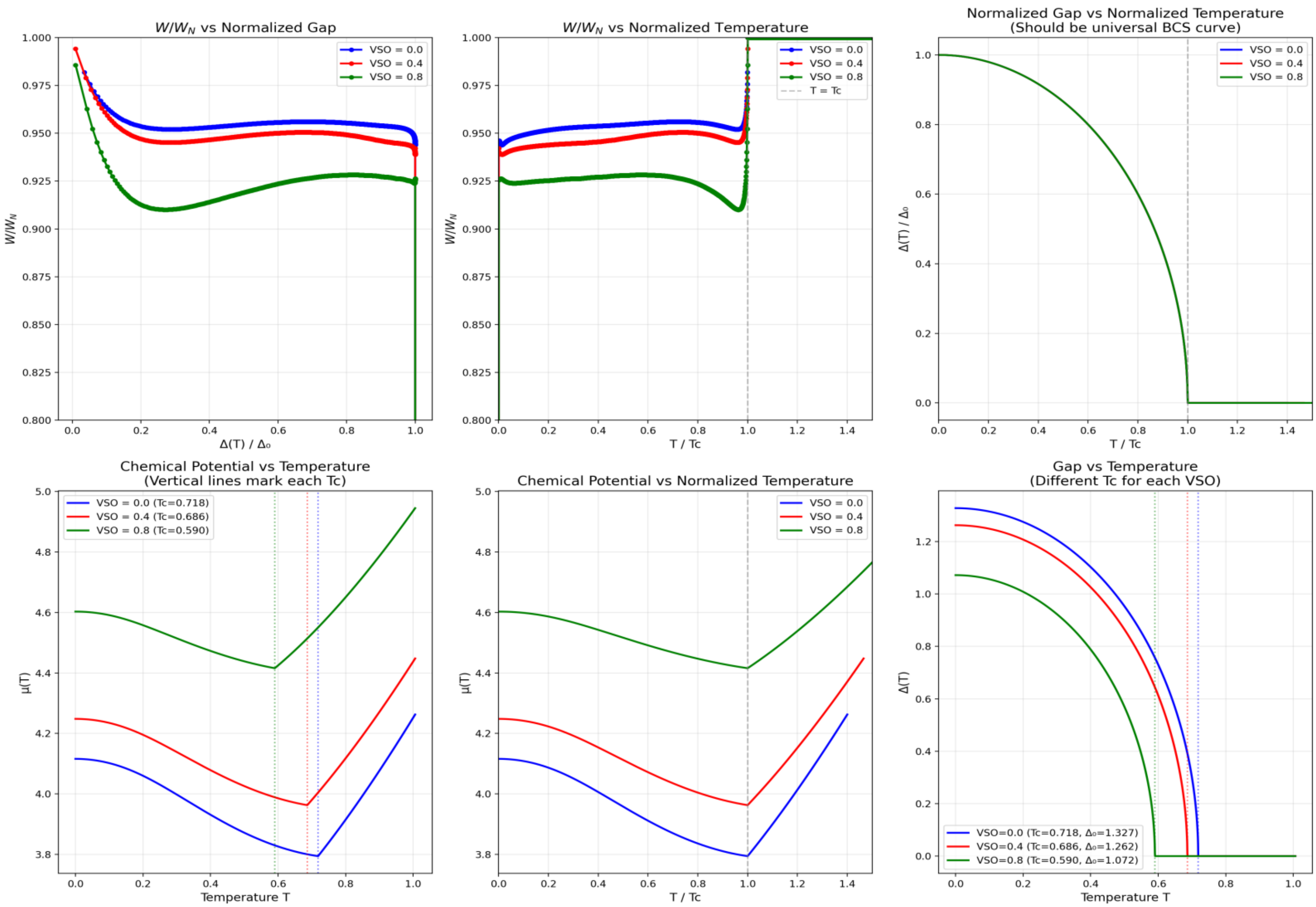

Ultrasonic Attenuation vs Spin-Orbit Coupling

$U$=95t , $n$=1.885, omega=0.7t

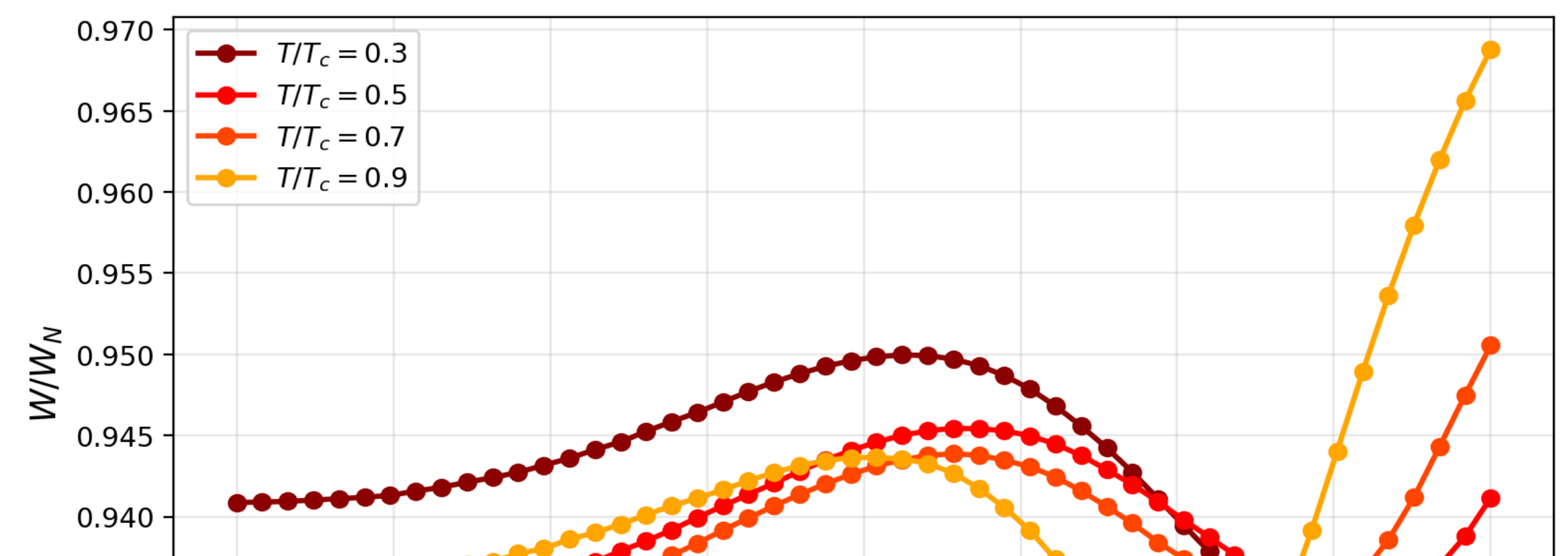

Ultrasonic Attenuation vs Spin-Orbit Coupling

$U$=125t , $n$=1.885, omega=0.7t

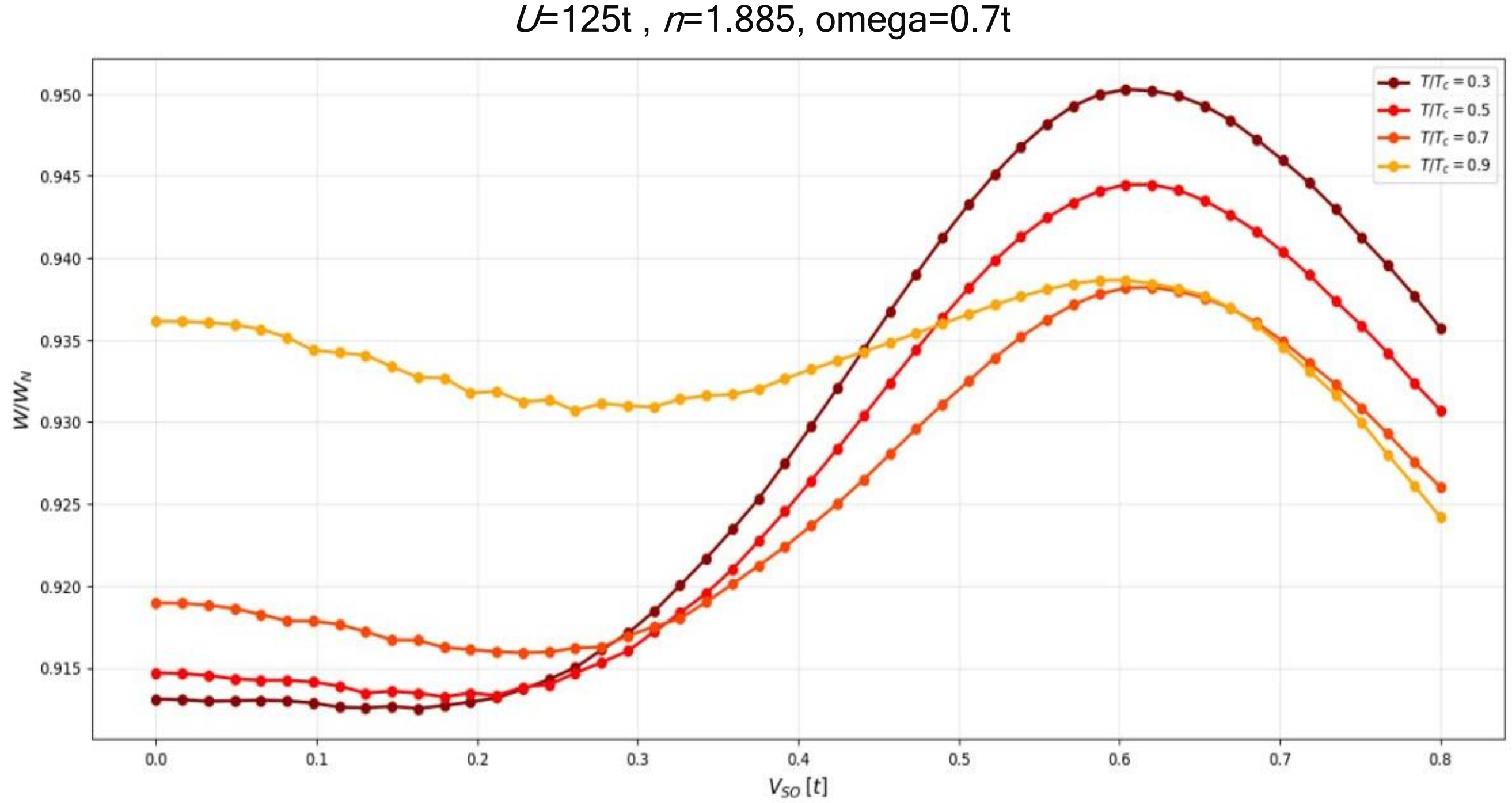

Ultrasonic Attenuation vs Spin-Orbit Coupling

$U$=-75t , $n$=1.885, omega=0.7t

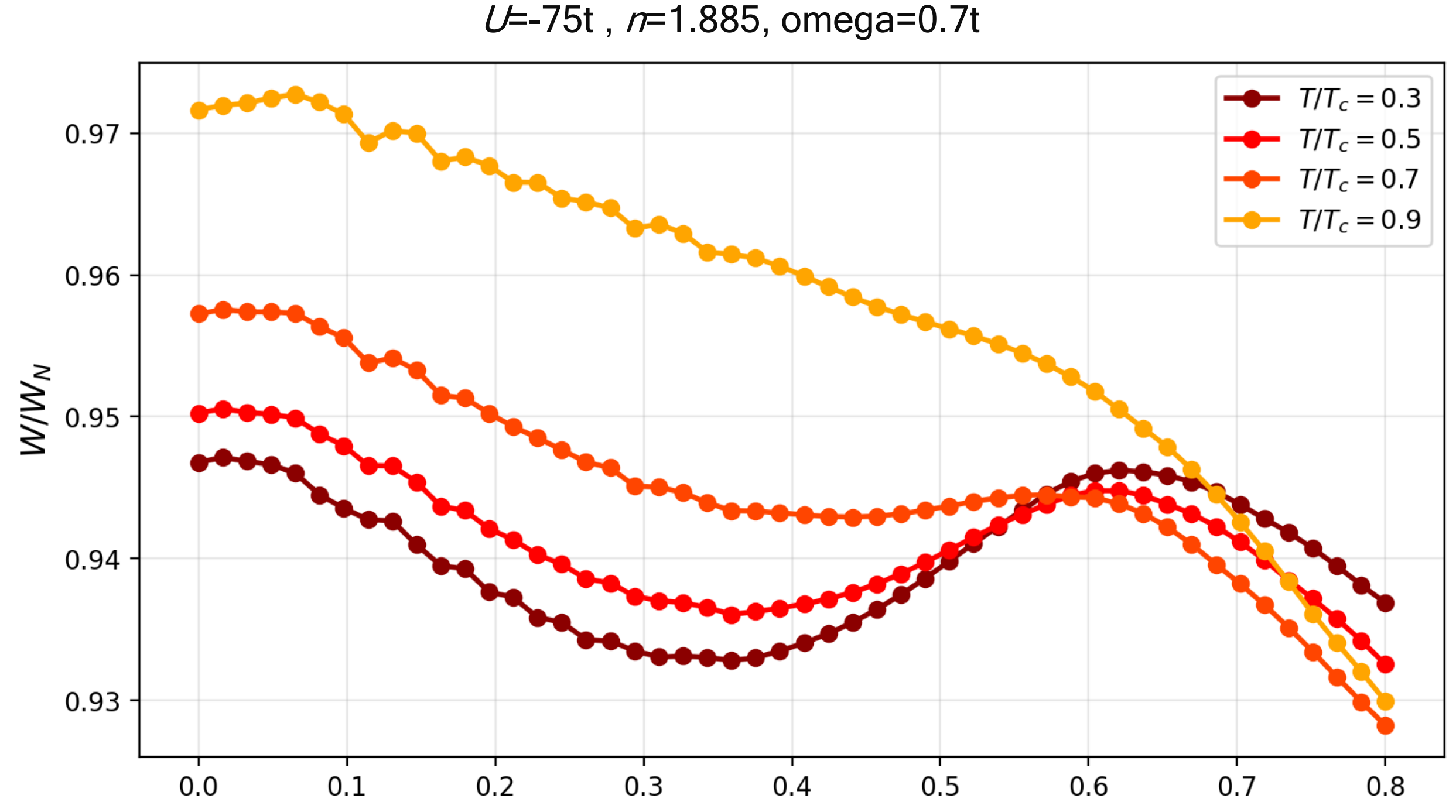

**Conclusion**

The provided study investigates the thermodynamic properties and superconducting response of a Hubbard model integrated with spin-orbit coupling ($V_{SO}$). Through a systematic analysis of the presented figures, it is observed that the superconducting state is highly sensitive to both thermal fluctuations and the intensity of spin-orbit interactions. The central observable, representing the normalized superfluid weight or absorbed power ($W/W_N$), demonstrates a consistent suppression as the temperature approaches the critical temperature ( $T_c$ ). This decline corresponds to the loss of superconducting coherence and the restoration of the normal-state behavior as the system crosses the transition. Crucially, the superconducting gap $\Delta(T)$ shows a monotonic decrease, vanishing precisely at $T_c$. When normalized as $\Delta(T)/\Delta_0$ and plotted against $T/T_c$. the data points collapse onto a universal BCS-like curve, indicating that the temperature dependence of the order parameter remains robustly conventional despite the specific Hubbard-model framework. A primary result of this work is the quantifiable enhancement of superconductivity driven by spin-orbit coupling. As the parameter $V_{SO}$ increases from $0.0$ to $0.4$ and finally to $0.8$ . There is a pronounced shift in the thermodynamic scales. These findings demonstrate that increasing spin-orbit coupling acts to strengthen pairing interactions and effectively stabilize the superconducting phase at higher temperatures. Furthermore, the chemical potential $\mu(T)$ exhibits a temperature-dependent shift, reflecting the underlying self-consistent electronic reorganization required to maintain the superconducting state across the transition. In summary, the figures illustrate that spin-orbit coupling significantly boosts superconducting performance, while the normalized thermodynamics adhere to universal, BCS-like behavior across all examined coupling strengths.

**References**

[1] S. B. Nam, "Microwave absorption in an unconventional superconductor ,(2006),Physical review letters, 96(11), 117001".

[2] F. S. a. I. V. T. Bergeret, "Spin-orbit coupling as a source of long-range triplet proximity effect in superconductor-ferromagnet hybrid structures." Physical Review B 89.13 (2014): 134517.".

[3] J. F. Annett, " Superconductivity, superfluids and condensates. Vol. 5. Oxford University Press, 2004".

[4] "M. Tinkham, "Introduction to Superconductivity Ch. 3 McGraw-Hill." Inc., New York 47 (1996)".".

[5] Y. A. a. E. I. R. Bychkov, "Oscillatory effects and the magnetic susceptibility of carriers in inversion layers." Journal of physics C: Solid state physics 17.33 (1984): 6039".

[6] F. S. A. F. V. a. K. B. E. Bergeret, "Odd triplet superconductivity and related phenomena in superconductor-ferromagnet structures." Reviews of modern physics 77.4 (2005): 1321".

[7] L. P. a. E. I. R. Gor'kov, "Superconducting 2D system with lifted spin degeneracy: mixed singlet-triplet state." Physical Review Letters 87.3 (2001): 037004.".

[8] G. M. Eliashberg, "FILM SUPERCONDUCTIVITY STIMULATED BY A HIGH-FREQUENCY FIELD. Inst. of Theoretical Physics, Moscow, 1970".

[9] S. B. (. Nam, ". Microwave absorption in an unconventional superconductor. Physical review letters, 96(11), 117001".

[10] M. e. a. Smidman, "Superconductivity and spin–orbit coupling in non-centrosymmetric materials: a review." Reports on Progress in Physics 80.3 (2017): 036501".

[11] M. e. a. Banerjee, "Observed quantization of anyonic heat flow." Nature 545.7652 (2017): 75-79.".

[12] H. K. Onnes, "The superconductivity of mercury." Comm. Phys. Lab. Univ. Leiden 122 (1911): 124.".

[13] J. L. N. C. a. J. R. S. Bardeen, "Theory of superconductivity." Physical review 108.5 (1957): 1175.".

[14] G. M. Eliashberg, "FILM SUPERCONDUCTIVITY STIMULATED BY A HIGH-FREQUENCY FIELD. Inst. of Theoretical Physics, Moscow, 1970.".

[15] C. C. a. J. R. K. Tsuei, "Pairing symmetry in cuprate superconductors." Reviews of Modern Physics 72.4 (2000): 969.".

[16] S. B. (. ]. Nam, " Microwave absorption in an unconventional superconductor. Physical review letters, 96(11), 117001".

[17] R. J. R. a. S. R. Micnas, "Superconductivity in narrow-band systems with local nonretarded attractive interactions." Reviews of Modern Physics 62.1 (1990): 113.".

[18] M. W. M. &. J. A. P. (. Alidoust, "Spin-superconducting fluctuations in mesoscopic systems. Physical Review B, 98(18), 184505".

[19] D. N. a. T. T. Basov, "Electrodynamics of high-T c superconductors." Reviews of modern physics 77.2 (2005): 721.".

[20] D. A. e. a. Bonn, "Evidence for rapid suppression of quasiparticle scattering below T c in YBa 2 Cu 3 O 7− δ." Physical review letters 68.15 (1992): 2390".

[21] "Micnas, R., J. Ranninger, and St Robaszkiewicz. "Superconductivity in narrow-band systems with local nonretarded attractive interactions." Reviews of Modern Physics 62.1 (1990): 113.".

[22] J. J. E. H. a. F. M. Hutchinson, ""Enhancement of superconducting T c due to the spin-orbit interaction." Physical Review B 97.18 (2018): 184513.".

[23] J. E. a. F. M. Hirsch, " "Superconducting state in an oxygen hole metal." Physical Review B 39.16 (1989): 11515.".

[24] "Ketterson, John Boyd, and Shengnian N. Song. Superconductivity. Cambridge university press, 1999.".